\documentclass[a4paper]{jpconf}
\usepackage{graphicx}
\usepackage{epstopdf}
\usepackage{subfigure,amsmath,amssymb,colordvi,times}
\usepackage[usenames]{color}
\usepackage{enumitem}
\def\agt{\,\hbox{\lower0.6ex\hbox{$\sim$}\llap{\raise0.6ex\hbox{$>$}}}\,}
\def\alt{\,\hbox{\lower0.6ex\hbox{$\sim$}\llap{\raise0.6ex\hbox{$<$}}}\,}
\begin{document}
\title{Many-body electron correlations in graphene}
\author{David Neilson$^{1}$, Andrea Perali$^{1}$ and Mohammad Zarenia$^{2}$}
\address{
$^{1}$ Dipartimenti di Fisica e di Farmacia, Universit\`a di Camerino, 62032 Camerino (MC), Italy\\
$^{2}$ Department of Physics, University of Antwerp,
B-2020 Antwerpen, Belgium}
\ead{david.neilson@unicam.it}

\begin{abstract}
The conduction electrons in graphene promise new opportunities to
access the region of strong many-body electron-electron correlations.
Extremely high quality, atomically flat two-dimensional electron sheets
and  quasi-one-dimensional electron nanoribbons with tuneable band
gaps that can be switched on by gates,  should exhibit new many-body
phenomena that have long been predicted for the regions of phase
space where the average Coulomb repulsions between electrons
dominate over their Fermi energies. In electron  nanoribbons a few
nanometres wide etched in monolayers of graphene, the quantum size
effects and the  van Hove singularities in their density of states further
act to enhance electron correlations. For graphene multilayers or
nanoribbons in a double unit electron-hole geometry, it is possible for
the many-body electron-hole correlations  to be made strong enough
to stabilise high-temperature electron-hole superfluidity.
\end{abstract}

\section{Introduction}

The Coulomb repulsion between conduction electrons in conventional
metals and semiconductors affects their properties only through the
relatively weak effects of linear screening and corrections to the values
of the Landau Fermi liquid parameters.
This is associated with the  high densities of conduction electrons
found in these
materials.  At such high densities, the Fermi energies are large or
comparable to the average electron-electron  interaction energies, and
so the electrons behave as a weakly-interacting system.  In such
systems, many-body electron correlations are so weak that they play only a
marginal role.

\subsection{New phenomena for strongly correlated conduction electrons}
If the density of conduction electrons could be lowered sufficiently to
make electron interactions dominate over Fermi energies, a wealth of
interesting new quantum phenomena  driven by strong many-body
electron correlations are predicted to appear.  These phenomena
include:
\begin{itemize}[label={}]
\item  Wigner crystal of electrons
\item  Charge density waves and other striped ground states
\item  Metal-insulator transition
\item  Quantum glass in the presence of defects
\item  Coherent superconductor and electron-hole superfluid quantum
states
\item  Crossover from BCS superconductivity to Bose-Einstein
condensation
\end{itemize}

\subsection{When are  many-body correlations expected to be important ?}
A two-dimensional (2D) electron layer is expected to have  correlations
that are stronger than the correlations in a corresponding three-
dimensional  system at the same density, because of the smaller kinetic
energy contributions.   The dimensionless parameter $r_s$ provides a
measure of when electron correlations will be important.  It is  defined
as  $r_s= \langle PE \rangle /  \langle KE \rangle $, and represents the
relative importance of the average kinetic energy at zero temperature,
$ \langle KE  \rangle = \langle \hbar^2 \nabla^2/2m^*\rangle
\simeq \langle \hbar^2/(m^* r_0^2)\rangle$, to the average strength
of the Coulomb potential energy from electron repulsion,
$\langle PE\rangle  = \langle e^2/\kappa r_0 \rangle$, where
$m^*$ is the electron effective mass and $\kappa$ the dielectric
constant of the background medium.  For electrons of density $n$ in a
2D layer, the average spacing between the electrons
$r_0=1/ \sqrt{\pi n}$.
In a 2D system with parabolic dispersion of the energy bands,
$r_s = r_0 / a_B^*$, where the effective Bohr radius for the system is
$a_B^*= \kappa\hbar^2/(e^2m^*)$.  Thus at sufficiently large $r_0$
(low electron densities), $r_s  \gg 1$, and the electron Coulomb
interactions will dominate over their Fermi energy.
	
\section{Strongly correlated conduction electron systems}

\subsection{2D Electron Liquid in Si MOSFETs and GaAs heterostructures}

\begin{figure}[ht!]
\centering
\includegraphics[height=50mm]{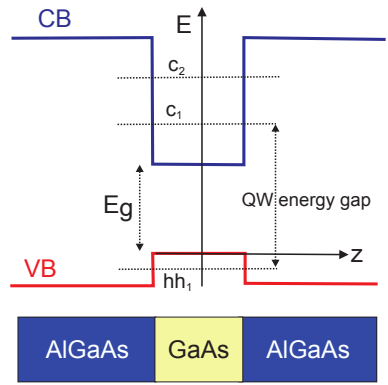}
\caption{Quantum well  in a GaAs heterostructure.\label{GaAsQW}}
\end{figure}

 In 2D systems, the new phenomena driven by strong correlations are
 predicted to appear at very low densities, generally only for $r_s\agt 10$ \cite{Swierkowski1991}.
 For many years  quasi-2D electron layers at the interfaces of Si
 MOSFET devices and in narrow quantum wells in GaAs
 heterostructures (Fig. \ref{GaAsQW}) have shown great promise as
 systems to access this  low density region \cite{Croxall2015}. However even in the highest
 quality semiconductors with extremely high mobilities, residual
 defects and impurities still block much access to this interesting
 region, since when the electron density is lowered by metal gates or by
 doping, the samples always  eventually become insulating.  This
 electron freeze-out occurs when the remaining conduction electrons
 can no longer screen residual charged impurities and they become
 trapped by the impurities.  The finite widths of the electron layers or
 quantum wells is a further challenge, as quantum wells must be wider
 than $10$-$15$ nm in order to retain their good conduction
 properties.  Finite widths mean weaker Coulomb interactions between
 electrons.

\subsection{Graphene}

In contrast to the quasi-2D electron layers in semiconductor systems, a
graphene sheet is atomically flat and hence strictly 2D.  There are no
finite-width effects to weaken the Coulomb interactions.
Levels of defects in graphene are extremely low so that electron freeze-out
should be postponed to much lower electron densities.

\subsubsection{Monolayer graphene}
\begin{figure}[ht!]
\centering
\includegraphics[height=35mm]{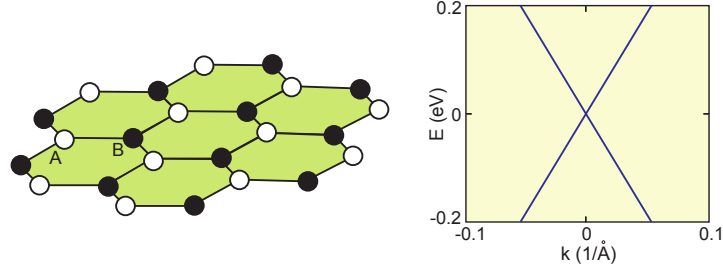}
\caption{Monolayer graphene: atomic structure and linear energy
bands at low energies. \label{MLG}}
\end{figure}

\ \\
However nature has not been kind.
Because the dispersion of the energy bands in  monolayer graphene is
linear at low energies,  $E_\pm(k) = \pm \hbar v_Fk$ (Fig.\ \ref{MLG}),
the Fermi energy  $E_F=\hbar v_F k_F=\hbar v_F \sqrt{\pi n}$ and  $
\langle KE \rangle$ depend only linearly on $r_0^{-1}$.     A result of
this is that  the $r_s= e^2/(\kappa \hbar v_F )$ in monolayer graphene
does not depend on the electron density.   The Fermi velocity
$v_F\simeq 10^6$ ms$^{-1}$ and the  dielectric constant of the
substrate is typically $\kappa \sim 3 $ \cite{Dean}, making the fixed
value of $r_s$  small, $r_s \sim 0.5$-$0.7$.  Such a small value for
$r_s$ implies that the correlations of the conduction electrons in
monolayer graphene will always remain weak \cite{sarma}.

\subsubsection{Bilayer graphene}
\begin{center}
\begin{figure}[ht!]
\centering
\includegraphics[height=60mm]{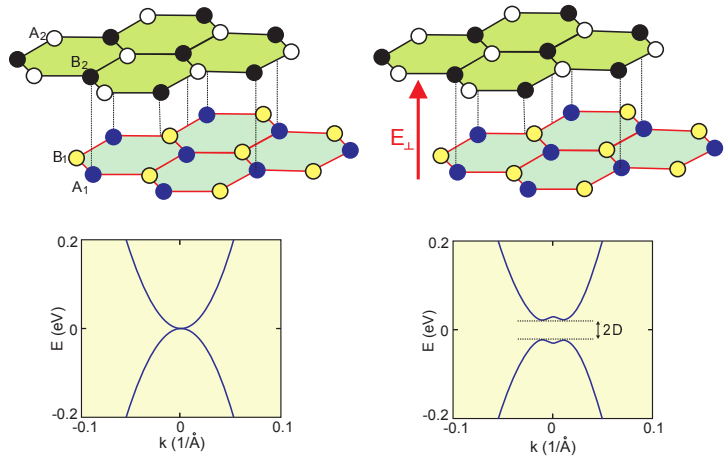}
\caption{Bilayer graphene: atomic structure and parabolic energy
bands at low energies.  With an external perpendicular electric field
$E_\perp$ there is an energy gap $2\Delta$ between the valence and conduction
bands. \label{BLG}}
\end{figure}
\end{center}

\ \\
It should, nevertheless, be possible to access the strongly correlated
region in graphene.  One work-around is to substitute a graphene
bilayer sheet in place of  the graphene monolayer.
A symmetrically biased graphene bilayer with AB stacking is a
semiconductor with parabolic dispersion of the energy bands
\cite{McCannFalkoPRL2006} (Fig.\ \ref{BLG}).  With parabolic
dispersion, the interaction parameter $r_s$ once again increases with
decreasing electron density as it does in a semiconductor system
\cite{Perali2013}.
Furthermore, in contrast to monolayer graphene, if a perpendicular
electric field is applied  across bilayer graphene using a metal gate, the
field will generate a tuneable energy gap between the conduction and
valence bands.

The  low density regime in bilayer graphene is dominated by disorder in
current samples, making a lower limit for the electron density of
$n\sim 10^{10}$ cm$^{-2}$, corresponding to a large maximum value
for $r_s=23$.  At very low densities a trigonal warping of the bands
could transform the parabolic bands into sets of Dirac-like linear bands
\cite{CastroNetoRMP2009}, but residual disorder will mask this effect,
and it could  be further reduced if necessary by applying an electric field
to open up an energy band gap.

In summary, extreme high quality, atomically flat bilayer graphene
sheets have tuneable electron densities and band gaps that should
permit the bilayers to readily access new quantum phenomena
predicted for strong electron correlations.

\subsubsection{Few-layer graphene}
\begin{figure}[ht]
\centering \vspace{0 mm}
\includegraphics[width=100mm]{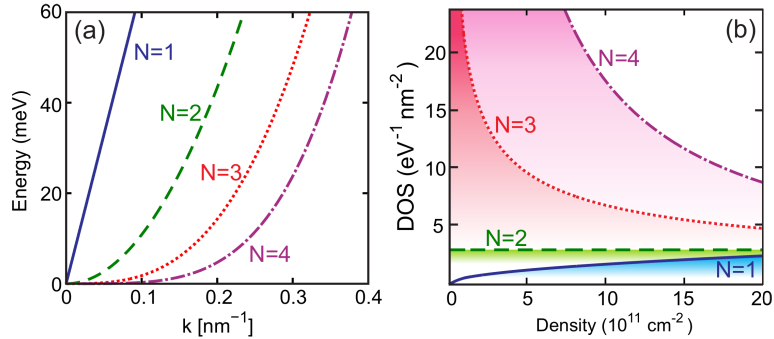}
\caption{  (a) Lowest positive energy band in monolayer ($N=1$),
bilayer ($N=2$), trilayer ($N=3$), and quadlayer ($N=4$) graphene.
(b) Comparison for $N=1$ to $N=4$ of the density of states at the
Fermi energy $DOS^{(N)}(E_F)$ as function of electron density
\cite{Zarenia}.  The van Hove power law singularities  as $E_F\rightarrow0$
lead to large enhancements of the DOS for trilayer and quadlayer graphene.}
\label{Bands_and_DOS}
\end{figure}

\ \\
Increasing the number of graphene layers in the sheet beyond bilayers
greatly enhances the density of states (DOS), and this projects the
sheets even more dramatically into the region of strong correlations at
accessible densities \cite{Zarenia}.   Electron graphene multilayers
should be able to  access regions of phase space with very strong
electron-electron interactions.  For the lowest energy band in ABC
stacked $N$-layer graphene,  the  energy dispersion of the conduction
band is given by \cite{min,katsnelson},
\begin{equation}
E^{(N)}(k)= \left\{(\hbar v_F)^{N}/t^{N-1}\right\} k^{N}\ ,
\label{E^N}
\end{equation}
where  $t\approx 400$ meV is the interlayer hopping term in few-layer
graphene.  Figure \ref{Bands_and_DOS}(a) shows $E^{(N)}(k)$ for
$N=1$ to $4$.   Using Eq.\ (\ref{E^N}) to determine the Fermi energy
$E_F$ at density $n$, we obtain for $N$-layer graphene,
\begin{equation}
r_s= \langle PE \rangle /  \langle KE \rangle = \left\{ \frac{e^2t^{N-1}}{\kappa (\hbar v_F)^N \sqrt{\pi^{N-1}}}\right\} \frac{1}{n^{(N-1)/2}}\ .
\end{equation}

A consequence of the different energy dispersions $E^{(N)}(k)$ is that
the dependence of the density of states on energy
changes dramatically with the number of layers $N$,
\begin{equation}
DOS^{(N)}(E)= \frac{d\Omega^{(N)}}{dE} = \frac{2\pi}{N}\frac{t^{2(N-1)/N}}{(\hbar v_F)^2}E^{(2/N)-1}\ ,
\end{equation}
where $\Omega^{(N)}(k)$ is the volume in  $k$-space of the $N$-layer
sheet.  Figure \ref{Bands_and_DOS}(b) shows the dependence of
$DOS^{(N)}(E_F)$  at the Fermi energy on electron density $n$.  For
monolayer graphene $DOS^{(1)}(E_F)$ depends linearly on $n$, for
bilayer graphene $DOS^{(2)}(E_F)$ is a constant, and for trilayer and
quadlayer graphene $DOS^{(N)}(E_F)$ decreases with increasing $n$.  Because
of the van Hove singularities as $E\rightarrow0$,  the
$DOS^{(3)}(E_F)$ and $DOS^{(4)}(E_F)$ for small densities are  much
larger than $DOS^{(1)}(E_F)$ and $DOS^{(2)}(E_F)$.  Eventually, at
very high densities lying well outside our  range of interest, the
$DOS^{(3)}(E_F)$ and $DOS^{(4)}(E_F)$ are smaller than
$DOS^{(1)}(E_F)$ and $DOS^{(2)}(E_F)$.

\begin{table}[ht]
\begin{center}
\centering
\begin{tabular}{c c c c c}
\hline\hline
Density (cm$^{-2}$) & monolayer   & bilayer & trilayer &  quadlayer     \\
\hline
$5\times 10^{12}$    &  $0.7$          & $1$      & $2$     &  $3$               \\
$1\times 10^{12}$    &  $0.7$          & $3$      &  $8$    &  $29$              \\
$5\times 10^{11}$    &  $0.7$          & $4$      &  $17$   & $83$              \\
$1\times 10^{11}$    &  $0.7$          & $8$      &  $86$   & $930$             \\
\hline
\hline
 \end{tabular}\
\label{table1}
\end{center}
\caption{Values of the interaction parameter $r_s$  for few-layer
graphene as a function of electron density.  Large values of $r_s\gg 1$
can lead to new phases driven by the strong electron correlations.}
\end{table}

Table I compares the values of $r_s$ for the typical electron densities
found in graphene sheets for $N$-layer graphene, for $N=1$
(monolayer) to $N=4$ (quadlayer).  The table shows that few-layer
graphene offers dramatic opportunities for producing extremely
strongly interacting electron systems at experimentally accessible
densities.

Experimental realisation of few layer graphene is  within the grasp of
current technology since few-layer graphene sheets can be fabricated
in large areas by both mechanical exfoliation
\cite{Ferrari2006,Zhang2005} and by chemical techniques
\cite{Berger2004,Shih2011,Mahanandia2014} from graphite with
controlled stacking order. References \cite{craciun,bao,mak} are
examples of experimental studies on electronic and transport
properties in trilayer graphene.
\subsubsection{Graphene nanoribbons}
\begin{figure}[ht]
\centering
\includegraphics[width=73mm]{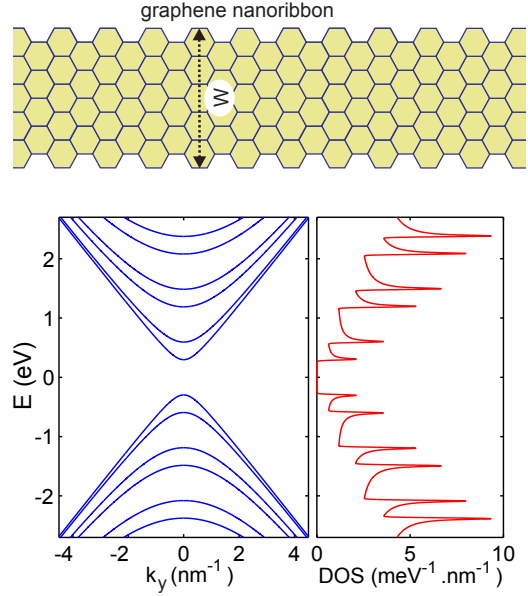}
\caption{
Top: armchair-edge terminated graphene nanoribbon of width $W$. Lower panels: lowest single-particle energy subbands $\epsilon_j(k_y), j=1,2,\dots$
for an armchair nanoribbon of width $W=2$ nm, and corresponding
density of states DOS$(E)$ in the nanoribbon \cite{Zarenia2015}. Van
Hove singularities are visible at the bottom of each subband.}
\label{Nanoribbon_bands}
\end{figure}

\ \\
Electrons can be confined in nanoribbons that are only
a few nanometres in width, etched on monolayers of graphene.
The nanoribbons can have multiple energy subbands that are occupied.
Their electronic properties  depend on the type of edge termination
\cite{Brey2007}.
We discuss here only  armchair-edge terminated graphene nanoribbons
(Fig.\ \ref{Nanoribbon_bands}), since, unlike for  zig-zag-edge terminated
nanoribbons, the multiple subbands of armchair
nanoribbons are parabolic around their minima and there is a
semiconductor-like energy gap between conduction and valence bands
(Fig.\ \ref{Nanoribbon_bands}).  Also, there is  no valley degeneracy.
These properties all act to  diminish the effect of Coulomb screening.
Uniform armchair graphene nanoribbons of widths much less than
$10$ nm have already been fabricated \cite{Cai}.

Electron correlations in armchair-edge terminated nanoribbons will be
further boosted  by the  quantum confinement of the electrons  along
the nanoribbons, and also by quantum size effects and van Hove
singularities in the quasi-one-dimensional density of states that are accessed if  the
Fermi energy is increased so it enters the bottom of a new subband.

Figure  \ref{Nanoribbon_bands} shows the single-particle
energy subbands $j$ in the continuum model, \\
$\epsilon_j(k_y)= (\sqrt{3}t a_0/2) \sqrt{k_y^2+k_j^2}$,  \
$j = 1,2,\dots$ for an armchair graphene nanoribbon of width
$W=2$ nm, where $t=2.7$ eV is the intralayer hopping energy
\cite{Brey} and $a_0=0.24$ nm is the graphene lattice constant.
We take the $y$-direction parallel to the nanoribbon, with the electrons
confined in the transverse $x$-direction.    The
quantised wave-number for the $j$-subband in the $x$-direction is \\
$k_j=\left[j\pi/W\right]-\left[4\pi/(3\sqrt{3}a_0)\right]$. The lower panels in Figure
\ref{Nanoribbon_bands} show the corresponding density of
states DOS$(E)$.  The van Hove singularities are clearly visible at the
bottom of each subband.

\subsection{Graphene in a periodic magnetic field}

A periodic magnetic field applied perpendicular to the graphene monolayers can preserve
the isotropic Dirac cones of the monolayer energy bands while reducing the slope of the
Dirac cones \cite{6,7,8,9}.  We represent the magnetic field perpendicular to the monolayer
as a one-dimensional array with period $2d$ of successive rectangular  magnetic barriers
and wells of height $B_z = \pm B$ and dimensionless width $d=d_B/\ell_B$, where
$\ell_B=\sqrt{\hbar c/eB} $ is the magnetic length.  Since the average of the magnetic flux
across a unit cell of the periodic field is  fixed at zero,  the main effect of the magnetic field
is to modify the electronic band structure of the graphene monolayers.  In this field,
the single-particle energy dispersion  of the monolayer graphene remains linear, but the
velocity $ \alpha_d v_F$ is less than the original Fermi velocity $v_F$, so
\begin{equation}
\epsilon({\bf k})=\pm \hbar( \alpha_d v_F) |{\bf k}| \left(1+ \delta({\bf k})\right)  \ .
\label{epsilonk}
\end{equation}
The non-linear correction term $\delta({\bf k})$ is small, with
$|\delta({\bf k})|\lesssim d_B^2k_x^2/6$.
The constant
$\alpha_d\le 1$,  representing the decrease in the Fermi velocity, depends on $d$ \cite{6,9},
\begin{eqnarray}
&\alpha_d\simeq 1-d^4/60 \ \ \ \ \  &d\ll 1\label{alpha_dsmall}\nonumber \\
&\alpha_d\simeq \frac{2d}{\sqrt{\pi}}e^{-d^2/4} \ \ \ \ \ &d\gg 1 \ .
\label{alpha_dlarge}
\end{eqnarray}

\begin{figure}[ht!]
\centering
\includegraphics[height=45mm]{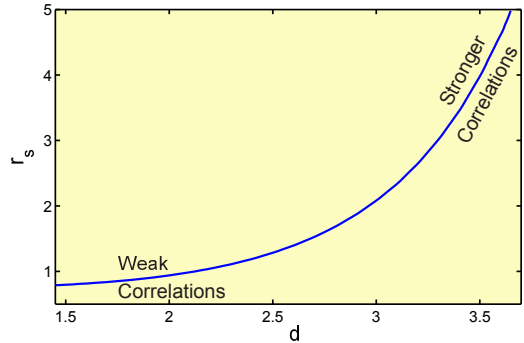}
\caption{
Variation of the interaction parameter $r_s$ for monolayer graphene in a perpendicular magnetic field of periodicity $2d=2d_B/\ell_B$ (see text).  Without the magnetic field, $r_s$ is a constant less than unity, and is independent of electron density.
}
\label{rs-alpha}
\end{figure}

Recalling  for graphene monolayers that the density-independent
interaction strength parameter is $r_s= e^2/(\kappa \hbar v_F )$, we see that an effect of the reduction of the Fermi velocity in Eq.\ \ref{epsilonk} is to increase the value of the $r_s$ parameter by  a factor $ \alpha_d^{-1}$.
Figure \ref{rs-alpha} shows that we can significantly increase $r_s$ by using the periodic magnetic field to tune $\alpha_d$ with $d$.

\section{Electron-hole superfluidity in graphene}

Two graphene monolayers of electrons and holes separated by a very
thin insulating barrier has  been proposed to observe an  electron-hole
superfluid \cite{SuperBLG24,MBSM2008,BMSMcomment}.
A  hexagonal boron nitride (hBN)  separation barrier as  thin as $1$ nm
can efficiently insulate the two  monolayers from each other
\cite{Gorbachev}.
However, theory suggests that because of strong screening of the
electron-hole pairing interaction, an electron-hole superfluid can
only occur when $r_s>2.3$ \cite{Lozovik2012}.  Experiments confirm
that the superfluid is not seen in two graphene monolayers
\cite{Gorbachev}, where we recall $r_s$ is fixed independent of density
at $r_s\sim 0.5$-$0.7$.

This poses a challenge as to whether new structures can be designed
and fabricated using atomically thin crystals, structures in which a
superfluid transition can be observed.
The graphene sheets discussed in Sections 2.2.2 to 2.2.4 can all access the
region $r_s\gg 2.3$, and hence they constitute promising
candidates to generate this elusive  electron-hole superfluid
\cite{Perali2013,Zarenia,Neilson2014,DellAnna}.

In  superfluid systems in graphene, it would be  straightforward to
access the BCS-BEC crossover and BEC regimes using the electric
potential on  metal gates and by tuning the sample parameters.  This
possibility opens up interesting new connections with the physics of
ultracold fermions and high-T$_c$ superconductors.

We have predicted the existence of electron-hole superfluidity in the
graphene structures described in Sections 2.2.2 to 2.2.4.  We find
mean field zero temperature superfluid gaps that are consistently
 large, of the order of several hundred Kelvin.  Unlike in three-dimensions,
 however, the superfluid phase transition temperature $T_c$
in a two-dimensional system is not related linearly to the magnitude of
its zero temperature superfluid gap.  For quantum condensates in
two-dimensions, an upper bound on the transition temperature is the
Kosterlitz-Thouless temperature \cite{KT1973},
$T_{KT} = (\pi/2)J(T_{KT})$, where $J(T)$ is the average kinetic energy
of the Cooper pairs.  At zero temperature, $J(0)$ is proportional to the
superfluid density, $J(0)=\rho_s(0)/2$, and, since in mean field
$\rho_s(T)$ depends very weakly on temperature for $T$ small
compared with the  zero-temperature superfluid gap, for this case
we may approximate  $J(T)$ by  $J(0)$ \cite{benfatto}.

We find the resulting $T_{KT}$ is typically of order $15$ K for the
electron-hole coupled graphene bilayers and the electron-hole coupled
graphene monolayers in a periodic magnetic field, for sheet
separations $\sim 2$ nm and typical experimental carrier
densities, $n\sim 10^{12}$ cm$^{-2}$.  Switching from coupled electron-hole
bilayers to the coupled electron-hole quadlayers for the same sheet separation,
typically doubles the $T_{KT}$.  These results indicate that electron-hole
superfluidity should be readily detectable in these graphene structures using
existing technology.

\section{Conclusions}

With graphene and related atomically thin crystals, we are at an
exciting many-body threshold to realise and exploit novel quantum
phases with tuneable properties.  High quality, atomically flat two-dimensional
electron graphene sheets and quasi-one-dimensional electron graphene nanoribbons
with tuneable electron densities and band gaps, should exhibit  novel
phenomena driven by strong many-body correlations.  These
phenomena are predicted when Coulomb repulsions between
electrons are dominant over their Fermi energies, and include the
Wigner crystal and charge density waves.  In addition, certain
configurations of double graphene sheets or nanoribbons,  one doped
with electrons and one with holes, should generate an electron-hole
superfluid state characterised by a very large superfluid energy gap and
relatively high transition temperature. 	

\ack
We thank Lucian Covaci and Fran\c{c}ois Peeters (University of Antwerp),
Alexander Hamilton (University of New South Wales),
and Luca Dell'Anna  (University of Padua) for  useful discussions.
We acknowledge support by the University of
Camerino FAR project CESEMN (DN and AP), the Flemish Science
Foundation (FWO-Vl) (MZ) and the University of Antwerp Research
Fund (BOF) (MZ). The authors thank colleagues involved in the
MultiSuper International Network (http://www.multisuper.org)
for exchange of ideas connected with this work.


\section*{References}
\begin{thebibliography}{99}

\bibitem{Swierkowski1991} \'{S}wierkowski L, Neilson D, and  Szyma\'{n}ski J 1991 Phys.\ Rev.\ Lett.\ {\bf 67}, 240;  Rapisarda Francesco and  Senatore Gaetano 1996 Aust.\ J.\ Phys.\ {\bf 49}, 161

\bibitem{Croxall2015} Zheng B, Croxall A F, Waldie J, Das Gupta K, Sfigakis F, Farrer I, Beere H E, Ritchie D A 
2015 ArXiv:1511.08701

\bibitem{Dean} Young A F, Dean  C R,  Meric  I,  Sorgenfrei  S,  Ren  H, Watanabe K,  Taniguchi T, Hone J, Shepard K  L and  Kim P 2012
Phys.\ Rev.\ B {\bf 85}, 235458

\bibitem{sarma} Dahal Hari P, Joglekar Yogesh N, Bedell Kevin S, Balatsky Alexander V 2006  Phys.\ Rev.\ B {\bf 74}, 233405;  Das Sarma S,  Adam Shaffique,  Hwang E  H and Rossi Enrico 2011 Rev.\ Mod.\ Phys.\ {\bf 83}, 407

\bibitem{McCannFalkoPRL2006} McCann E and  Fal'ko V I 2006 Phys.\ Rev.\ Lett.\ {\bf 96}, 086805

\bibitem{Perali2013} Perali A, Neilson D and  Hamilton A R 2013 Phys.\ Rev.\ Lett.\ {\bf 110}, 146803

\bibitem{CastroNetoRMP2009} Castro Neto A H,  Guinea F, Peres  N M R, Novoselov K  S and Geim A K 2009 Rev.\ Mod.\ Phys.\ {\bf 81}, 109

\bibitem{Zarenia} Zarenia M, Perali A,  Neilson D and  Peeters F  M 2014 Sci.\ Reports {\bf 4}, 7319

\bibitem{min} Min H and  MacDonald A  H 2008 Phys.\ Rev.\ B {\bf 77}, 155416

\bibitem{katsnelson} Katsnelson, M  I  2012 {\it Graphene: Carbon in two dimensions} (New York:  Cambridge University Press)

\bibitem{Ferrari2006}  Ferrari A  C, Meyer J  C,  Scardaci V,  Casiraghi C,  Lazzeri M,  Mauri F,  Piscanec S,  Jiang D,   Novoselov K  S,   Roth S and   Geim A  K 2006 Phys.\ Rev.\ Lett.\ {\bf 97}, 187401

\bibitem{Zhang2005}  Zhang Y,   Small J  P,   Pontius W  V and  Kim Philip 2005 Appl.\ Phys.\ Lett.\ {\bf 86}, 073104

\bibitem{Berger2004}   Berger C,   Song Z,   Li T,  Li X, Ogbazghi  A  Y,   Feng R,   Dai Z,   Marchenkov A  N,   Conrad E  H,   First P  N  and  de Heer W  A 2004 J. Phys.\ Chem.\ B {\bf 108}, 19912

\bibitem{Shih2011} Shih Chih-Jen,   Vijayaraghavan A,   Krishnan R,  Sharma R,	 Han Jae-Hee,  Ham Moon-Ho,   Jin Zh,   Lin Sh,   Paulus G  L C,   Reuel N  F,   Wang Q  H,   Blankschtein D and   Strano M  S 2011 Nature Nanotechnology {\bf 6}, 439

\bibitem{Mahanandia2014}   Mahanandia P,   Simon F,    Heinrich G and   Nanda K  K 2014 Chem.\ Commun.\ {\bf 50}, 4613

\bibitem{craciun}   Craciun M F,   Russo S,   Yamamoto M,   Oostinga J B,   Morpurgo A F and   Tarucha S 2009 Nature Nanotech.\ {\bf 4}, 383

\bibitem{bao}  Bao W,   Jing L,   Velas J,   Aykol M,  Cronin S  B,   Smirnov D,   Koshino M,   McCann E,   Bockrath M and   Lau C  N 2011 Nature Phys.\ {\bf 7}, 948

\bibitem{mak} Mak  Kin Fai,  Shan Jie and   Heinz Tony F 2010 Phys.\ Rev.\ Lett.\ {\bf 104}, 176404

\bibitem{Brey2007} Brey L and Fertig H A 2007 Phys.\ Rev.\ B {\bf 75}, 125434

\bibitem{Cai}   Tapaszt\'{o} L,   Dobrik G,   Lambin P and   Bir\'{o} L P 2008 Nature Nanotechnology {\bf 3}, 397

\bibitem{Brey}   Brey L and   Fertig H A 2006 Phys.\ Rev.\ B {\bf 73}, 235411

\bibitem{Zarenia2015} Zarenia M, Perali A,  Peeters F  M and Neilson D 2015 ArXiv:1601.06942

\bibitem{6}  Dell'Anna L and   De Martino A 2009 Phys.\ Rev.\ B {\bf 79}, 045420; {\it ibid.} 2009 {\bf 80}, 089901(E)

\bibitem{7}    Snyman I 2009 Phys.\ Rev.\ B {\bf 80}, 054303

\bibitem{8}  Tan L  Z,   Park C -H and  Louie  S  G 2010 Phys.\ Rev.\ B {\bf 81}, 195426

\bibitem{9} Dell'Anna L and   De Martino A 2011 Phys.\ Rev.\ B {\bf 83}, 155449

\bibitem{SuperBLG24}  Zhang C -H and   Joglekar Y N 2008 Phys.\ Rev.\ B {\bf 77}, 233405;
Lozovik Yu E  and Sokolik A A 2008 JETP Lett.\ {\bf 87}, 55;
2010 Eur.\ Phys.\ J.\ B {\bf 73}, 195;
Mink M P,  Stoof H T C,   Duine R A and  MacDonald  A H 2011 Phys.\ Rev.\ B {\bf 84}, 155409

\bibitem{MBSM2008}  Min Hongki,  Bistritzer Rafi,  Su Jung-Jung and MacDonald A  H  2008 Phys.\ Rev.\ B {\bf 78}, 121401(R)

\bibitem{BMSMcomment}   Bistritzer R,   Min H,  Su  J -J and   MacDonald A H 2008
ArXiv:cond-mat/0810.0331v1

\bibitem{Gorbachev} Gorbachev R V,   Geim A K,   Katsnelson M I,   Novoselov K S,   Tudorovskiy T, Grigorieva I V,  MacDonald A H,   Watanabe K,   Taniguchi T and  Ponomarenko L A 2012
Nat.\ Phys.\ {\bf 8}, 896


\bibitem{Lozovik2012} Lozovik Yu E,   Ogarkov S L and Sokolik A A 2012
Phys.\ Rev.\ B {\bf 86}, 045429

\bibitem{Neilson2014}   Neilson D,  Perali A and  Hamilton A R 2014 Phys.\ Rev.\ B \ {\bf 89}, 060502(R)

\bibitem{DellAnna} Dell'Anna  L,  Perali A, L  Covaci  L and  Neilson D 2015 Phys.\ Rev.\ B {\bf 92} 220502(R) 

\bibitem{KT1973} Kosterlitz J M and Thouless D J 1973 J.\ Phys.\ C: Sol.\ State Phys., {\bf 6}, 1181

\bibitem{benfatto} Benfatto L, Capone M, Caprara S, Castellani C, and Di Castro C 2008 Phys. Rev. B {\bf 78}, 140502(R); Benfatto L and Sharapov S G 2006 Low Temp.\ Physics {\bf 32}, 533

\end{thebibliography}
\end{document}